\documentclass[twocolumn,nofootinbib,showpacs,pre,final]{revtex4}
\begin{document}
 
\title{Aggregation According to Classical Kinetics---From Nucleation to Coarsening}
\author{Yossi Farjoun}
 \email{yfarjoun@math.mit.edu}
\thanks{Corresponding author}
\affiliation{Department of Mathematics, Massachusetts Institute of Technology}
\author{John C. Neu}%
\email{neu@math.berkeley.edu}
\affiliation{Department of Mathematics, University of California, Berkeley}
\date{\today}

\numberwithin{equation}{section}        
\renewcommand{\theequation}{\arabic{section}.\arabic{equation}}
\newcommand{\E}{{e}}

\newcommand{\RR}{{\cal R}}
\newcommand{\DD}{{\cal D}}
\newcommand{\ud}[1]{{\,{d}#1}}

\newcommand{\tsvalue}{{8.9037}}

\begin{abstract}
We solve the standard Lifshitz-Slyozov (LS) model with conservation of total particles
in the limit of small super-saturation.  
The new element is an effective initial condition that follows from the initial exhaustion of nucleation as described in a previous paper \cite{FN:Creation:08}. 
The effective initial condition is characterized by a narrow distribution of cluster-sizes, all much larger than critical.
In the subsequent solution, one of the LS similarity solutions emerges as the long-time limit, as expected.
But our solution tells more. 
In particular, there is a ``growth'' era prior to what is usually called ``coarsening.''
During ``growth'' the clusters  (all of nearly the same size much larger than critical) eventually exhaust the super-saturation (the exhaustion of \emph{nucleation} in the previous era results from only a small decrease in super-saturation). 
This allows the critical size to catch up to the clusters, and the traditional ``coarsening'' begins: Subcritical clusters dissolve and fuel the growth of the remaining super-critical clusters.
Our analysis tracks the evolution of cluster sizes during growth and coarsening by complimentary use of asymptotic and numerical methods. 
We establish characteristic times and cluster sizes associated with growth and coarsening from physical parameters and the initial super-saturation.
The emerging distribution is discontinuous at the largest cluster size, and thus selects the discontinuous LS similarity solution as the long-time limit.
There are strong indications that the
smooth similarity solution proposed in the original LS paper emerges on a, yet longer, ``late-coarsening'' time-scale.
\end{abstract}


\maketitle
\section*{Introduction}
\label{sec:intro}
We analyze the growth of clusters in a monomer bath with conserved total monomer density in the small super-saturation limit. 
The description of late stage coarsening according to the classic Lifshitz-Slyozov (LS) theory \cite{LS61} is well known in the literature of aggregation:
The number of monomers in the largest clusters increases linearly in time, and the density of clusters shrinks as the smaller clusters dissolve back into monomers. 
But how does it all begin, starting from pure monomer at ``time zero''?
In a previous paper \cite{FN:Creation:08} the authors  predict the
cluster size distribution that emerges from the nucleation process,
and its long-term limit. 
In the current paper, the solution of the LS equations, with that
 long-term limit used as an effective initial condition, is tracked all the way to late stage coarsening.
In this way, we obtain a ``big picture'' of the whole aggregation process, based on classical modeling ideas due to Becker-D\"oring \cite{BD35}, Zeldovich \cite{ZELD43}, and Lifshitz-Slyozov \cite{LS61}.
\begin{figure*}[ht]

\begin{center}
\begin{minipage}{12cm}
\centerline{
\resizebox{12cm}{!}{\includegraphics{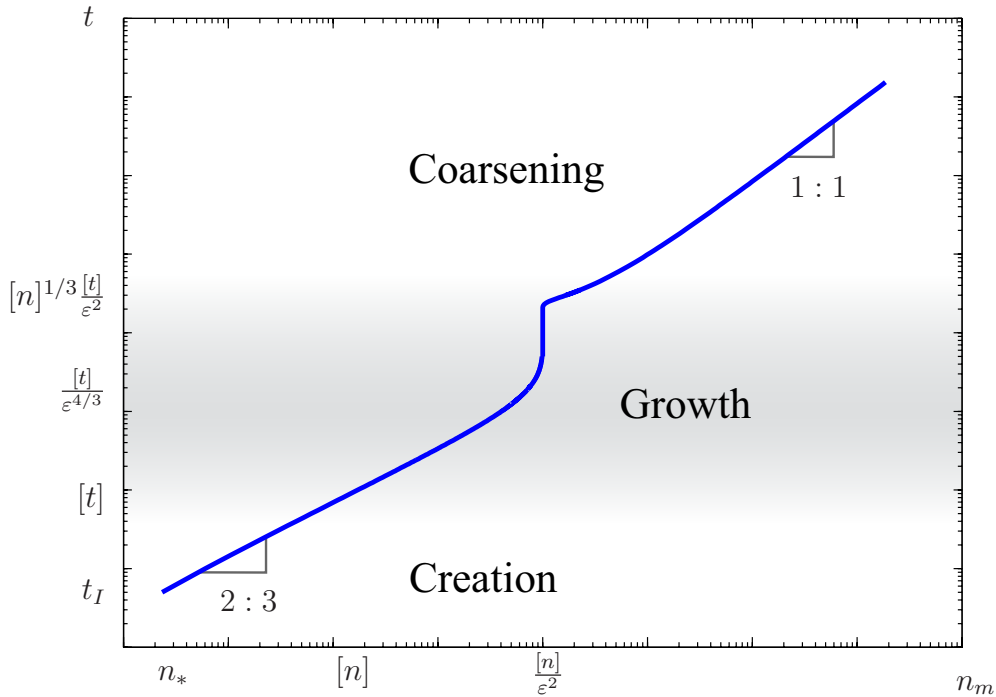}}}
\caption[An overview graph showing the size of the largest cluster versus time.]{The graph of time vs.\ maximal cluster-size in logarithmic  scale. The graph shows two obvious regimes (creation and coarsening)
  separated by a ``kink'' in the graph (growth and ``pre-coarsening'')
The scales $n_*$, $[n]$ and $\sig^3 [n]/R
  \eps^2$ of cluster size are marked in the plot.
}
 \label{fig:whole_picture}
\end{minipage}
\end{center}
\end{figure*}

Here is the summary using Fig.~\eqref{fig:whole_picture} as a visual guide.
The horizontal axis is the largest cluster size, $n_{m}$, the vertical is time $t$, both with logarithmic scales.
This graph of $t$ vs.\ $n_{m}$ is based on the quantitative solution of the complete model.
The plane is divided into horizontal time-slices, ``Creation'', ``Growth'', and ``Coarsening.''
The characteristic time $[t]$ to exhaust nucleation is exponentially large in the initial free 
energy barrier $G_{*}$ against nucleation, with $[t]\propto\exp\brk{\tfrac25 G_{*}/k_{B}T}$. The timescale $[t]$ is the thickness of the creation time-slice.
In this time, the supersaturation undergoes only a small relative decrease and the initial clusters continue rapid growth. 
For diffusion limited growth in (nearly) constant super-saturation, the number of monomers $n(t)$ in a cluster grows at a rate proportional to cluster radius, so $\dot n\propto n^{1/3}$ and it follows that $n(t)\propto t^{3/2}$. 
In particular, the asymptotic line in the creation time-slice of Fig.~\ref{fig:whole_picture} has a $2:3$ slope consistent with $n_{m}^{2/3}\propto\frac{t}{[t]}$. 
In this way we see that the characteristic size $[n]$ of clusters during the creation era is proportional to $[t]^{3/2}\propto \exp\brk{\tfrac35 G_{*}/k_BT}$.
The width of the cluster size distribution grows more slowly, like $\brk{t/[t]}^{1/2}$. 
Hence, the relative width of the cluster size distribution becomes small during the tail of the creation era, $\frac{t}{[t]}\goto\infty$.
The actual profile of this narrow distribution is determined from the time history of the nucleation rate per unit volume $j(t)$, derived in \cite{FN:Creation:08}.
 
In the next time-slice, labeled ``Growth'', the nearly homogeneous population of rapidly expanding clusters seriously depletes the super-saturation.
This depletion causes their rapid growth to stop when their (common) cluster size reaches $\frac{[n]}{\eps^{2}}$.
Here, $\eps$ with $0<\eps\ll1$ is the initial super-saturation.
In Fig.~\ref{fig:whole_picture}, this arrested growth is represented by the nearly vertical segment.
The characteristic time which measures the thickness of the growth time-slice is $\frac{[t]}{\eps^{4/3}}$.

The next era, \emph{coarsening}, begins when the critical cluster
size, much smaller than the characteristic size during
creation and growth,
``catches up'' with the clusters' size.
Clusters smaller than critical shrink, and the monomers they
shed are taken up by the larger, growing clusters. 
Thus, the distribution widens.
The characteristic size of the clusters during the coarsening era remains the
same as it was during the growth era, but the timescale is much
longer: $[n]^\third[t]/\eps^2$.

In \emph{late stage} coarsening, $t\gg[n]^{\frac13}[t]/\eps^{2}$, the cluster size distribution asymptotes to a similarity solution of the LS equations, and we finally reach the stage when $n_{m}$ is linear in time.
Indeed, the asymptotic line in the coarsening time-slice of Fig.~\ref{fig:whole_picture} has the characteristic $1:1$ slope.

Thus concludes the ``brief history'' of aggregation according to the classical ideas of BD, Zeldovich, and LS.
We highlight some collateral results.
First regarding time and size scales: 
By introducing a physical initial condition that represents the initial nucleation process, we ultimately determine the characteristic time to reach coarsening and the characteristic cluster size, as functions of the physical parameters and initial supersaturation. 
In particular the time to reach coarsening, $[t]_{\text{c}}=[n]^{\third}[t]/\eps^{2}$, is exponentially large in the initial free energy barrier $G_{*}$, even relative to the time $[t]$ of the creation era.
An actual physical time-scale \emph{cannot} result from the LS theory alone.
Indeed, the famous similarity solutions result precisely from scale invariance.

This brings us to a peculiar detail:
The late-stage coarsening similarity solution that is selected by our solution of the LS equations is discontinuous at the largest cluster size.
It is widely believed that the physically correct similarity solution is the smooth, $C^{\infty}$, one.
In the discussion section we propose that during an additional era following coarsening, the distribution evolves further and tends to the smooth $C^{\infty}$ similarity solution.

The organization of the paper is as follows: 
In Section~\ref{sec:creation}, we review the origins of the cluster size distribution during the tail of the creation era, as set forth in \cite{FN:Creation:08}.
This, of course, is the effective initial condition for the growth era, treated in Section~\ref{sec:growth}.
Section~\ref{sec:coarsening} treats the coarsening era and its asymptotic matching with the tail of the growth era.
Here  there is an additional twist:
The coarsening era is evolved numerically, whereas the effective
initial condition inherited from the growth era comes from an analytic
solution.
The switching from analytic to numerical solution is controlled as a
function of the numerical resolution, so we have a de-facto
``analytic-numerical matching.''
One corollary of this expanded sense of matching is the analytic determination
of a time delay for the onset of coarsening, proportional to $\log
\oneover{\eps}$.

In Section~\ref{sec:similarity} we re-derive the family of similarity solutions using our notation. 
The discontinuous member of this family is determined and the convergence of the numerical solution to it is verified.
\section{The Physical Model and Effective Initial Conditions}
\label{sec:creation}
In the classic Lifshitz-Slyozov (LS) theory, the number of monomers $n=n(t)$ in a cluster satisfies the ODE of diffusion limited growth:
\begin{equation}
\dot n = \DD(\eta n^{\third}-\sig),  \quad \DD=\brk{3(4\pi)^{2}}^{\third} D v^{\third}f_{s}.
\label{eq:ODE:n}
\end{equation}
Here, $\eta$ is the chemical potential of monomers in the bath (in units of $k_{B}T$) relative to monomers in the bulk of clusters.
When the monomer density $f_{1}$ approaches the \emph{saturation
  density} $f_{s}$, for which the monomer bath would be in equilibrium
with an ``infinite'' cluster, we have the asymptotically linear relation
\begin{equation}
\eta = \tfrac{f_{1}-f_{s}}{f_{s}}.
\label{eq:eta}
\end{equation}
In \eqref{eq:ODE:n}, $\sig$ is the dimensionless surface tension constant so that the interfacial free energy of a cluster with $n$ monomers is $\frac32n^{\frac23}\sig k_{B}T$.
In the definition of the rate constant $\DD$, $D$ denotes the diffusivity of monomers in the bath and $v$ is the monomer volume inside clusters.

The state variable of the LS equations is the cluster-size distribution $r(n,\,t)$, so that the density of clusters with size $n$ between $n_{1}$ and $n_{2}$ is $\int_{n_{1}}^{n_{2}}r(n,\,t)\ud{n}.$
There are two basic equations: 
First,  the convection PDE 
\begin{equation}
\partial_t r + \partial_n\BRK{\DD\brk{\eta n^\third -\sig}r}=0, \text{ for } n>0,
\label{eq:nucleation:PDE:full}
\end{equation}
represents transport of clusters in the space of their size $n$ by the diffusion limited growth ``velocity'' in \eqref{eq:ODE:n}.
Second, the conservation of monomers couples the value of the super-saturation, $\eta$ and the solution. 
The conservation of monomer is expressed approximately by
\begin{equation}
f= (1+\eta)f_{s}+\int_{0}^{\infty}n\,r(n,\,t)\ud{n}.
\label{eq:integral}
\end{equation}
Here, the total monomer density $f$, a constant in time, is the sum of
monomer density $f_{1}=(1+\eta)f_{s}$ (from \eqref{eq:eta}) in the
bath, and the the density of monomers in clusters is approximated by the integral.

In the convection PDE \eqref{eq:nucleation:PDE:full}, $\sig$ is positive, so characteristics in the (n,t) plane are \emph{absorbed} by the $n-$axis.
Hence, the $n-$axis is a ``sink'', representing the complete dissolution of subcritical clusters. 
This is consistent with the assumption that creation of new clusters by fluctuation over the critical size is negligible during the ``growth'' and  ``coarsening'' eras. 
In a previous paper \cite{FN:Creation:08} we derive scaling units $[t],\,[r],\,[n]$ of time $t$, cluster size $n$ and cluster size density $r$ that characterize the creation era. 
It is convenient to express the characteristic scales of the growth and coarsening eras as multiples of these creation era scales.
Hence, we carry out a preliminary non-dimensionalization of (\ref{eq:nucleation:PDE:full}, \ref{eq:integral}) based on $[t],\,[r]$, and $[n]$.
The unit of chemical potential $\eta$ is $[\eta]=\eps$, the initial value of chemical potential in the pure monomer bath, before nucleation.
The dimensionless equations are
\begin{align}
\partial_tr&+\partial_n\BRK{n^\third r-s}=0,\quad \text{ in } n>0,\label{eq:PDE:scaled}\\
\eta &= 1-\frac{\eps^2}{\sig^3}\int_0^\infty n\, r\,dn.
\label{eq:conservation:scaled}
\end{align}
In \eqref{eq:PDE:scaled}, $s$ is the scaled surface tension, exponentially small as $\eps\goto0$ defined in appendix~\ref{app:scaling} in \eqref{eq:s}.
Equations (\ref{eq:PDE:scaled}, \ref{eq:conservation:scaled}) are
solved for $r(n,\,t)$ subject to an effective initial condition that arises from asymptotic matching with the creation era.
In the previous paper we showed that at a range of time $t$, after
nucleation is exhausted, but before the effects of growth change the
super-saturation significantly, $r(n,\,t)$ is asymptotic to a narrow
distribution is approximated by  
\begin{equation}
  r(n,\,t)= \left\{
\begin{aligned}
&N^{-\third}j \brk{\frac{N-n}{N^{1/3}}}, &0< N-n < 5 N^{\third}\\
&0, & \text{otherwise}.
\end{aligned}
\right.
\label{eq:nucleation:solution}
\end{equation}
Here, $n=N(t)$ is the size of the largest cluster, approximated by 
\begin{equation}
N(t)\sim\brk{\tfrac23 t}^{\frac32} (\text{for } t=\O(1)).
\label{eq:nucleation:growth}
\end{equation}

\begin{figure}[tb]
  \centering
  \resizebox{8cm}{!}{\includegraphics{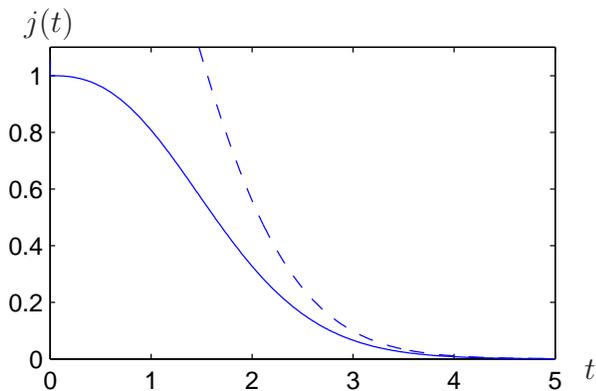}}
  \caption{The solution of Eq.~\eqref{eq:integral} is the rate of
    production of  new clusters $j=\E^{\delta\eta(t)}$ during the
    nucleation era. The tail of the flux rate decays
    super-exponentially in $t$. The dashed line is the asymptotic
    result found in \cite{FN:Creation:08}.}
  \label{fig:flux}
\end{figure}

The function $j(t)$ is the dimensionless nucleation rate whose graph is shown in Fig.~\ref{fig:flux}.
In our previous paper \cite{FN:Creation:08} it is shown that $j(t)$ satisfies the integral equation 
\begin{equation}
\log j(t)=-\int_{0}^{t} \brk{\tfrac23(t-\tau)}^{\frac32}j(\tau)\ud{\tau}.\label{eq:nucleation:flux}
\end{equation}
Its solution, $j(t)$, decays to zero faster than exponential as $t\goto\infty$.
For $t\ge5$, $j(t)$ is a negligible fraction of its initial value, so
we truncate the support of $j(t)$ to $0<t<5$ for simplicity of presentation.
This explains the upper limit $5N^{\third}$ in \eqref{eq:nucleation:solution}.
The reduction in monomer density due to aggregation during the creation era is denoted $R$ and is calculated from~$j$:
\newcommand{\Rd}{{1.7117}}
\begin{equation}
R = \int_0^\infty j(\tau)\ud{\tau} \approx \Rd.\label{eq:total:density:nuc}
\end{equation}
The value $R\approx\Rd$ is, of course, a scaled density. 
To get a physical density one needs to multiply it by $[n] [r]$, with $[n]$ and $[r]$ given by (\ref{eq:scale:n}, \ref{eq:scale:r}).
\section{Growth Era}
\label{sec:growth}
During the growth era, the cluster distribution is still approximated by 
\eqref{eq:nucleation:solution}, but the growth of the largest cluster-size $N(t)$ slows relative to the $t^{3/2}$ growth law \eqref{eq:nucleation:growth} due to the depletion of supersaturation. 
Here is a brief summary of the argument.
In the convection PDE \eqref{eq:PDE:scaled}, the component $\eta n^{1/3}$ of convection velocity is much greater than one, so the scaled surface tension $s$ is asymptotically negligible.
The convection PDE thus reduces asymptotically to 
\begin{equation}
\partial_tr+\partial_n\BRK{n^\third r}=0,\quad \text{ in }
n>0.\label{eq:PDE:reduced}
\end{equation}
The corresponding \emph{physical} idea is that most of the clusters are much larger than critical.
It follows from \eqref{eq:PDE:reduced} that $n^{1/3}r(n,\,t)$ is constant along
characteristics that satisfy
\begin{equation}
\dot n = \eta n^\third. 
\label{eq:char:n}
\end{equation}
In \eqref{eq:char:n}, $\eta=\eta(t)$ decreases from (near) 1 in the
beginning of the growth era to (near) 0 at the end in a manner
consistent with the conservation identity
\eqref{eq:conservation:scaled}. 
We see that the characteristics determined by \eqref{eq:char:n} are \emph{continuations} of the creation era characteristics, carrying the \emph{same} values of $n^{1/3}r$.

\begin{figure}[tb]
  \flushright
  \resizebox{8cm}{!}{\includegraphics{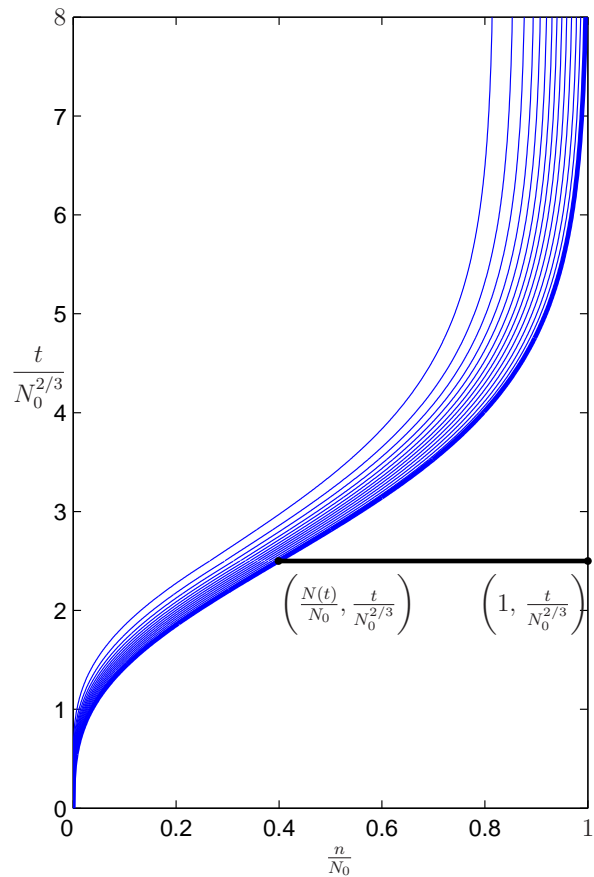}}
  \caption{The ``world-lines'' of clusters created at the origin. The
    density of the lines corresponds to the density of clusters at
    each point. The length of the horizontal line from  $(N(t),\,t)$ to $(N_0,\,t)$ (in units of $N_0$) is the supersaturation $\eta$ (in units of $\eps$.)}
\label{fig:char:log}
\end{figure}

This indicates a very simple construction of the asymptotic solution for $r(n,\,t)$ during the growth era. 
The details are in Appendix \ref{app:growth}.
In summary,  $r(n,\,t)$ is concentrated in a narrow front near the
largest cluster size $N$, and there approximation
\eqref{eq:nucleation:solution} applies.
What changes is the evolution of $N(t)$, now described by the ODE
\begin{equation}
\dot N = N^\third \brk{1-\frac{N}{N_0}}, \quad N_0=\frac{\sig^3}{\eps^2 R}.
\label{eq:N_0}
\end{equation}
Here, $1-\frac{N}{N_0}$ is the value of $\eta(t)$ consistent with the conservation identity \eqref{eq:conservation:scaled}.
The solution to ODE \eqref{eq:N_0} with $N(0)=0$ and $N(t)>0$ for $t>0$ is given implicitly by
\begin{equation}
\frac{t}{N_0^\frac23} = \sum_{j=0}^2 r_j
\log\brk{1+r_j\brk{\frac{N}{N_0}}^\third},\label{eq:ODE:sol:growth}
\end{equation}
where $r_j$ are the cube roots of $-1$: $r_0=\E^{i\frac{\pi}3}, r_1=-1,
r_2=\E^{-i\frac{\pi}3}$.
Figure~\ref{fig:char:log} shows this solution as a ``world line'' in the $(n, t)$ plane (dark line).
The shaded area represents the front where $r(n,\,t)$ is concentrated.
In the limit $1\ll t\ll N_0^{2/3}$, \eqref{eq:ODE:sol:growth} reduces to 
$t\sim \frac32 N^{2/3}$,
in agreement with results \eqref{eq:nucleation:growth} from the creation era.

In the opposite limit $t\gg N_0^{2/3}$, which corresponds to the \emph{tail-end}
of the growth era, $N(t)$ asymptotes (exponentially) to the constant value
$N_0=\sig^3/\eps^2 R$. 
The size distribution asymptotes to 
\begin{equation}
r\sim N_0^{-\third} j\brk{\frac{N_0-n}{N_0^\third}}, \label{eq:r:longtime:growth}
\end{equation}
independent of time.
It is still narrow, with its support is concentrated in an
interval of $n$ with $0<N_0-n=\O(N_0^\third)\ll N_0$.

Why does the size distribution ``stop dead in its tracks''?
In Fig.~\ref{fig:char:log}, the length of the horizontal line segment from
$\brk{N(t)/N_0,\,t}$ to $(1,\,t)$ represents the super-saturation $\eta$ (in
units of $\eps$) at time $t$.
It asymptotes to zero for $t\gg N_0^{2/3}$, and the \emph{truncated}
convection velocity $\eta n^{1/3}d$ vanishes with it.
The clusters  ``use up'' the super-saturation that  fuels their growth. 

In summary, during the growth era, the clusters grow in a relatively
narrow distribution until they reach a maximal cluster size
$n=N_0$ (in units of $[n]$). 
The width of the distribution is proportional to $N_0^{1/3}$.
The time-scale of the era is $N_0^{2/3}$ (in units of $[t]$), and
roughly 10 of these time-units are needed for the narrow, stationary
distribution to be established, as seen in Fig.~\ref{fig:char:log}. 
The growth of the clusters is fueled by the supersaturation, which
vanishes in an asymptotic sense.
\section{Coarsening Era}
\label{sec:coarsening}
The apparent ``road-block'' to further growth is not the end of the
aggregation story. 
The growth era asymptotics are not uniformly valid as $t/N_0^{2/3}
\goto \infty$. 
As $\eta$ decreases, the exponentially small component $s$ in the full
convection velocity $u=\eta n^\third - s$ in
\eqref{eq:nucleation:PDE:full} gains influence until it balances the (now small) $\eta n^\third.$
The \emph{critical size} $n_*\equiv \brk{s/\eta}^3$, where $u=0$,
``catches up'' with the average cluster size and is now near $N_0$.
Clusters smaller than the critical size $n_*$ shrink,
shedding monomers and fueling the continued growth of the clusters
larger than $n_*$. 
The classic process called \emph{coarsening} has begun.
The characteristic time of coarsening, to be determined shortly, is
exponentially longer than the characteristic time $\Brk{t} N_0^{2/3}$ of the
growth era.
During coarsening, the distribution widens and eventually  fills the
whole range of cluster sizes from the (growing) maximal size down to zero. 
The tail of the coarsening era is characterized by convergence to
one of the self-similar distributions predicted by Lifshitz and Slyozov.

\subsection{Coarsening era scaling}
Relative scaling units%
\footnote{Since the convection PDE \eqref{eq:nucleation:PDE:full} and conservation identity \eqref{eq:conservation:scaled}
are non-dimensionalized using nucleation era units $[t],\,[n]$ from
(\ref{eq:scale:t}, \ref{eq:scale:n}) for $t$ and $n$, and $\eps$ is the unit for $\eta$, dominant
balances in these equations provide scaling units \emph{relative} to
those of the nucleation era.
For instance the characteristic cluster size relative to $[n]$ is
$N_0$ in \eqref{eq:N_0}, and the actual unit
of $n$ is $N_0[n]$.}
of time $t$ and super-saturation $\eta$ follow
from the balance of all three terms in the convection velocity $u=\eta
n^\third -s$ in \eqref{eq:nucleation:PDE:full}.
The balance between $u$ and $s$ yields $N_0/s$ as the
relative unit of time, while balancing $\eta n^{1/3}$ and $s$
gives $s/N_0^{1/3}$ as the relative unit of $\eta$.
The relative unit $R/N_0$ of $r$ follows from the balance of the two
terms in the RHS of the conservation identity \eqref{eq:conservation:scaled}.
The relative and absolute units of $n,\,t,\,\eta$, and $r$ are
summarized in the scaling table:
\begin{center}
\begin{tabular}{l*{4}{|>{$}c<{$}}}
Variable\hspace*{2mm}&n&t&\eta&r\\
\hline
Relative Unit&N_0=\frac{\sig^3}{\eps^2 R}&\frac{N_0}{s}&\frac{s}{N_0^{1/3}}& \frac{R}{N_0}\\
Absolute Unit&N_0[n]&\frac{N_0}{s}[t]&\frac{\eps s}{N_0^{1/3}}& \frac{R}{N_0}[r]\\
\end{tabular}
\end{center}
The largest cluster size $N(t)$ satisfies ODE \eqref{eq:ODE:n}.
In the new units this ODE reads
\begin{equation}
\dot N = \eta N^\third -1.\label{eq:ODE:N:coarsen}
\end{equation}
The scaled PDE \eqref{eq:nucleation:PDE:full} and conservation
identity \eqref{eq:conservation:scaled} are now
\begin{align}
\partial_t r &+ \partial_n\BRK{\brk{\eta
    n^\third-1}r}=0,\label{eq:PDE:coarsen}
\intertext{in $0<n<N$, and }
\frac{s}{N_0^{1/3}}\eta &= 1- \int_0^N n\,r\ud{n}.\label{eq:int:nr:full}
\end{align}

\subsection{The determination of the supersaturation}
In the analysis of the creation and growth eras, the conservation
identity explicitly determines $\eta$ from $r(n,\,t)$.
In the coarsening era this straightforward approach fails:
By assuming $\eps\ll1$ we also get $s/N_0^{1/3}\ll1$, hence the leading order
approximation of \eqref{eq:int:nr:full} is 
\begin{equation}
 \int_0^N n\,r\ud{n}=1.\label{eq:int:nr}
\end{equation}
The term containing $\eta$ disappears.
Physically, most of the available monomers are contained in clusters
and the super-saturation is vanishingly small.
To extract $\eta$ from $r(n,\,t)$ we differentiate \eqref{eq:int:nr} with respect to $t$:
\begin{equation}
\dot N \, r(N,t)+\int_0^N n\, \partial_t r\ud{n} =0. \label{eq:dt:int:nr}
\end{equation}
Next, we substitute $\dot N$ from \eqref{eq:ODE:N:coarsen}, and
$\partial_t r$ from the
convection PDE \eqref{eq:PDE:coarsen} into \eqref{eq:dt:int:nr} and
integrate by parts. 
After some some algebra we find that $\eta$ can be expressed as
\begin{equation}
\eta =\frac{\int_0^N r\ud{n}}{\int_0^N n^\third r\ud{n}}.
\label{eq:eta:explicit}
\end{equation}
In summary, $r(n,\,t)$ in $0<n<N$ satisfies the integro-differential
equation, consisting of the convection PDE \eqref{eq:PDE:coarsen} with $\eta$ as in
\eqref{eq:eta:explicit} and $N$ as in \eqref{eq:ODE:N:coarsen}.
An effective initial condition is determined by asymptotic matching
with the tail of the growth era.
At $n=0$ the convection velocity is negative, so a boundary
condition there is not required. 

\subsection{Changing variables}
The growth of the largest cluster size $N(t)$ with time implies that
 PDE \eqref{eq:PDE:coarsen} has to be solved on a growing interval of $n$. 
We simplify the numerical solution by using the following change of
variables first:
\begin{equation}
x\equiv\frac{n}{N},\qquad q(x,\,t)\equiv N r(N x,t).\label{eq:new:q:x}
\end{equation}
The idea should be clear;
The normalized cluster size $x$ ranges in the \emph{fixed} interval
$(0,1)$ and $q$ is the distribution of cluster sizes in $x-$space.
We multiply $r$ by $N$ so that $q\ud{x} = r\ud{n}$.
The convection PDE \eqref{eq:PDE:coarsen} for $r(n,\,t)$ transforms
into an convection PDE for $q(x,\,t)$,
\begin{equation}
\partial_t q + \partial_x\BRK{w\,q}=0,\label{eq:PDE:coarsen:2}
\end{equation}
in $0< x <1$. 
Here, $w$ is the convection velocity in $x$ space,
\begin{equation}
w = \oneover{N}\brk{\eta N^\third (x^\third-x)+(x-1)}.\label{eq:ad:vel:coarsen}
\end{equation}
Boundary conditions are not required, since $w$ vanishes at $x=1$ and is negative at $x=0$.
Equation \eqref{eq:eta:explicit} translates into a
  functional dependence of $\eta$ upon $N$ and moments of $q$,
  \begin{equation}
  N^\third\eta =\frac{\int_0^1 q\ud{x}}{\int_0^1 x^\third q\ud{x}}. \label{eq:eta:xq:1}
   \end{equation}
The largest cluster size $N$ is easily determined from the conservation identity
\eqref{eq:int:nr} written in terms of $q$ and $x$:
\begin{equation}
\oneover{N}=\int_0^1 x\,q\ud{x}.\label{eq:int:xq}
\end{equation}
In summary, both $\eta$ and $N$ are found explicitly from $q(\cdot,\,t)$ on the interval $(0,\,1)$, and this makes \eqref{eq:PDE:coarsen:2} an explicit integro-differential evolution equation for $q$.
It is convenient to introduce the moments of $q(x,\,t)$ (themselves
functions of time):
\begin{equation}
  M_0 \equiv \int_0^1 q\ud{x}, \quad
  M_\third \equiv \int_0^1 x^\third q\ud{x}, \quad
  M_1 \equiv \int_0^1x\,q\ud{x}. \label{eq:Moments}
 \end{equation}
Then \eqref{eq:eta:xq:1} and \eqref{eq:int:xq} become 
\begin{equation}
N^\third\eta = \frac{M_0}{M_\third},\quad 
N = \oneover{M_1},\label{eq:N:M}
\end{equation}
and the convection velocity $w$ can be written as
\begin{equation}
w=M_1\brk{\frac{M_0}{M_\third} (x^\third-x) + (x-1)}.\label{eq:ad:vel:coarsen:2}
 \end{equation}
\subsection{Initial conditions and early widening}
The $t\goto0$ limit of the coarsening solution for $r(n,\,t)$ should
match distribution \eqref{eq:r:longtime:growth}, which characterizes
the tail of the growth era.
Hence, we have the effective initial condition
\begin{equation}
  q(x,\,0) = N_0^\frac23 j\brk{N_0^\frac23\brk{1-x}}. \label{eq:init:cond:coarsen}
\end{equation}
The RHS in \eqref{eq:r:longtime:growth} is translated into $x,\,q$
variables.
Since $N_0\gg1$, this initial distribution is a
tall spike of height $N_0^{2/3}$ concentrated in a narrow interval of $x-$values near $x=1$:
$0\le1-x\le\O(N_0^{-2/3})$.
The initial condition for the largest cluster size $N$ (in coarsening
units) is $N(0)=1$.

To our knowledge, the integro-differential evolution equation for
$q(x,\,t)$ does not admit an analytic solution, so a numerical solution
is sought.
From a numerical point of view, the tall, narrow initial condition
\eqref{eq:init:cond:coarsen} is not desirable for two reasons:
First, it is narrow, with width proportional to $\eps^{\frac43}$, and
thus resolving it numerically would be difficult (for $\eps\ll1$).
Second, this initial condition depends on $\eps$ via the dependence on
$N_0$, thus for every $\eps$ we would need to run the computation again.
Some preliminary asymptotics fixes both issues and supplies us with a
global solution:
As long as the distribution remains a narrow spike near $x=1$, and thus the
three moments---$M_0,\  M_\third$, and $M_1$---are all near 1, the convection velocity $w$ in
\eqref{eq:ad:vel:coarsen:2} can be approximated by
\begin{equation}
w\sim x^\third-1 = \tfrac{1}{3} (x-1)+ \O(x-1)^2,
\label{eq:asymptotic:vel:coarsen}
\end{equation}
near $x=1$.
The convection PDE \eqref{eq:PDE:coarsen:2} with $w$ replaced by
its linearization \eqref{eq:asymptotic:vel:coarsen} can be solved analytically:
The ``early'' evolution of $q(x,\,t)$ based upon the linearized
convection velocity \eqref{eq:asymptotic:vel:coarsen} is given by the
widening distribution:
\begin{equation}
  q(x,\,t) = N_0^\frac23 \E^{-t/3} j\brk{N_0^\frac23
    \E^{-t/3}\brk{1-x}}.
\label{eq:q:asymptotic:coarsen}
\end{equation}
This asymptotic distribution matches the effective initial condition
\eqref{eq:init:cond:coarsen} for $t=0$, and remains valid as long as
the ``$x-$width'' remains small, $N_0^{-\frac23} \E^{t/3}\ll1$.
\subsection{Time-shift and the numerical solution}
The strategy now is as follows: 
First, we assume that our numerical PDE solver accurately resolves a distribution of width $\delta$ with
$N_0^{-2/3}\ll\delta\ll1$.
From the $\eps-$dependent initial condition
\eqref{eq:init:cond:coarsen},  we evolve $q(x,\,t)$ according to the
asymptotic solution \eqref{eq:q:asymptotic:coarsen} 
until the ``$x-$width'' $N_0^{-2/3} \E^{t/3}$ achieves the value
$\delta$. 
This happens at time
\begin{equation}
t = 2\log N_0 +3\log \tfrac1{\delta}.\label{eq:time:num:start}
\end{equation}
The numerical solver takes over for times greater than $t$ in
\eqref{eq:time:num:start}.
The width $\delta$ is chosen so that it is much larger than the numerical
discretization of $x$, so that the solution can be resolved, yet much
smaller than 1 so that the analytic solution remains valid.

\begin{figure}[ht!]
 \centerline{
\resizebox{\columnwidth}{!}{\includegraphics{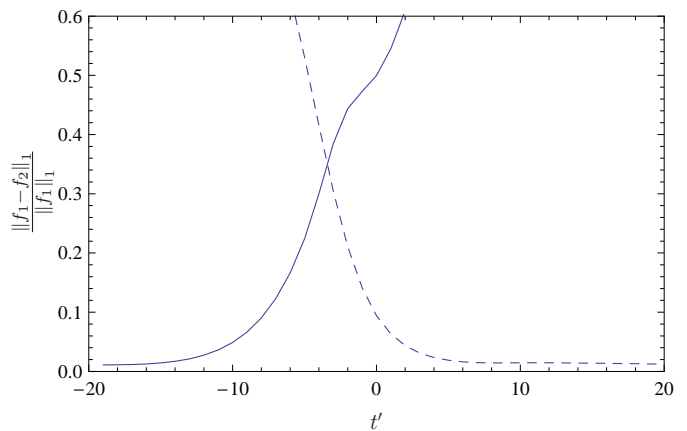}}}
\caption{The normalized distance between the numerical solution and the asymptotic solution given by \ref{eq:q:asymptotic:coarsen} (solid),
  and  the distance between the numerical solution and the discontinuous similarity solution given by \eqref{eq:sim:sol:final} (dashed).}
\label{fig:NumericalMatching:f}

\end{figure}

It is convenient to absorb the $\eps-$dependent component $2\log N_0$
in \eqref{eq:time:num:start} by shifting the origin of time.
The shifted time is
\begin{equation}
t' = t - 2\log N_0\label{eq:t':t}
\end{equation}
and the numerical solver is turned on at \emph{shifted} time $t'=3\log
\oneover{\delta}$, with the effective initial condition
\begin{equation}
  q(x,\,t') = \tfrac1{\delta} j\brk{\tfrac1\delta \brk{1-x}},
\label{eq:effective:IC}
\end{equation}
in $0<x<1$.
As desired, the time-shift produces an  $\eps-$\emph{independent} initial
condition for the numerical solver, and thus an $\eps-$independent
numerical solution.
For a wide range of $\delta$'s in $N_0^{-2/3}\ll\delta\ll1$, the
numerical solution at \emph{fixed} $t'$ should be close to the
asymptotic solution. 
We use this later (see Fig.~\ref{fig:NumericalMatching:f}) to convince
ourselves of the numerical solver's acceptable performance.
The details of the numerical solution are spelled out in the Section~\ref{sec:methods}.

\begin{figure}[ht!]

 \centerline{
\resizebox{9cm}{!}{\includegraphics{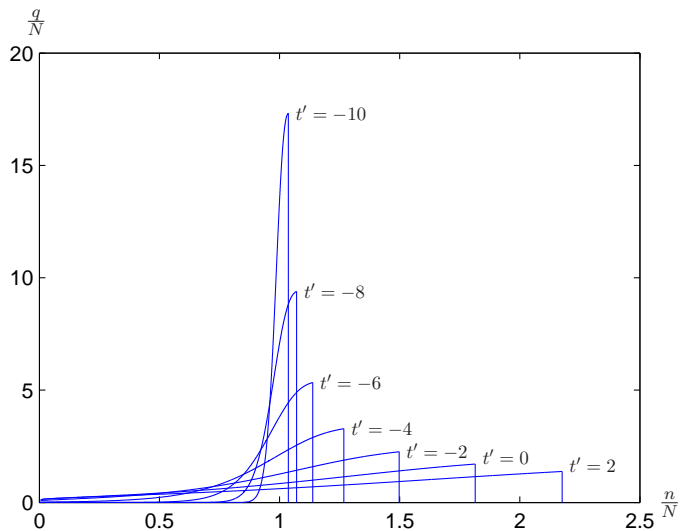}}}
\caption[The cluster size distribution during the coarsening era]{The numerical solution at various times as found using
  \texttt{clawpack}. Displayed are snapshots from 
  $t'=-10$ to $t'=2$. The solution continues to evolve after $t'=2$, as it converges to the similarity solution. Prior to
$t'=-10$, the solution is described by the (analytic) asymptotic solution.}
\label{fig:coarsening:1}
\end{figure}

After finding $q(x,t')$ numerically, we reconstruct $r(n,\,t)$ using
\eqref{eq:new:q:x}:
\begin{equation}
r(n,t) = \tfrac1{N} q\brk{\tfrac{n}{N},t-2\log N_0}.\label{eq:reconst:coarsen}
\end{equation}
For $t<3\log\delta +2\log N_0$, we use the asymptotic expression \eqref{eq:q:asymptotic:coarsen}
for $q$, and for $t>3\log\delta +2\log N_0$, we use the numerical solution.
Figure~\ref{fig:coarsening:1} shows the \emph{numerical solution} for the coarsening era at various
values of the \emph{shifted} time $t'$. 

\begin{figure}[ht!]
 \centerline{
\resizebox{\columnwidth}{!}{\includegraphics{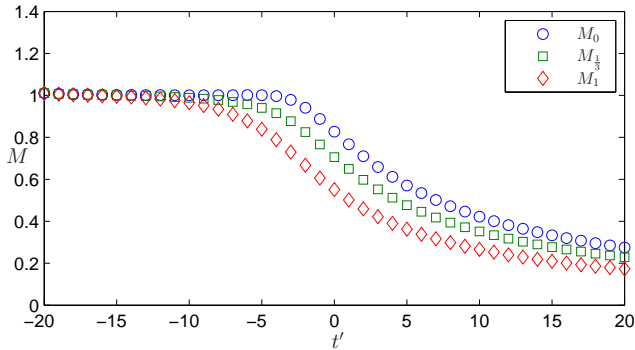}}}
\caption{The three moments $M_0$, $M_\third$, and $M_1$ calculated for
the numerical solution. Around $t'=-10$, their distance from 
 1 is noticeable and  the asymptotic solution is no
longer valid. Around $t'=-5$, $M_0$ departs from $1$ clusters start dissolved at $x=0$.}
\label{fig:NumericalMatching:M}
\end{figure}

The coarsening era solution exhibits three phases: widening,
transition, and similarity solution (also called \emph{late stage coarsening}).
During the initial widening, the support of the distribution has
not yet reached $x=0$, and the fraction of clusters which have
dissolved completely is negligible (See
Fig.~\ref{fig:NumericalMatching:M}).
The widening is accurately described by the asymptotic solution
\eqref{eq:q:asymptotic:coarsen}.
During the transition, the support of $q$ reaches down to $x=0$ and the
smaller clusters start dissolving, so the total density of clusters 
decreases. 
To resolve this part of the solution the numerical solver is
required. 
The solution is shown in Fig.~\ref{fig:coarsening:1}.
During the ``tail'' of coarsening we can observe the convergence of the
distribution to a specific similarity solutions of the LS model.

The three phases of the coarsening era can be seen in 
Fig.~\ref{fig:NumericalMatching:f} which shows the (normalized) distance\footnote{We use  $\frac{\int_0^1\Abs{n(x)-a(x)}
    \ud x}{\int_0^1\Abs{a(x)}\ud x}$ to measure the distance between a numerical solution
  $n(x)$ and an asymptotic solution $a(x)$.
  The normalization is used because the similarity solution decays to 0 as $t\goto\infty$ and thus a simple norm might give an impression of convergence when there is none.
}
between the
numerical solution and the asymptotic solution (the solid line), and
between the numerical solution and the discontinuous similarity
solution (the dashed line).  
Initially, the numerical solution agrees with the asymptotic
solution and the normalized distance is negligible.
Afterward, in the transition, non-linear effects and the non-zero
width of the distribution cause a widening ``rift'' between the 
numerical solution and asymptotic one. 
These non-linear effects also drive the numerical solution towards the
similarity solution (which is described in greater detail
below), until eventually, the numerical solution is almost
indistinguishable from one of the similarity solutions.

\section{Similarity Solutions}
\label{sec:similarity}

We briefly review the two-parameter family of LS similarity solutions. 

The convection PDE (\ref{eq:PDE:coarsen:2}--\ref{eq:int:xq}) admits a  separation of variables solution:
\begin{equation}
q(x,t')=c(t')\,P(x).\label{eq:ss:ansatz}
\end{equation}
We start with the temporal part $c(t')$:
In the ODE \eqref{eq:ODE:N:coarsen} for $N(t)$, substitute $N=1/{M_1}$ and
$N^{1/3}\eta={M_0}/{M_{1/3}}$ as follows from the two equations
in \eqref{eq:N:M}.
We get
\begin{equation}
\dot M_1 = M_1^2\cdot\brk{1-\tfrac{M_0}{M_\third}} \label{eq:M:dot}
\end{equation} 
and  equations (\ref{eq:Moments}, \ref{eq:N:M}, \ref{eq:ss:ansatz}, \ref{eq:M:dot}) imply an ODE for $c(t')$:
\begin{equation}
  \dot c = -c^2F\cdot(\mu-1)\label{eq:c_dot},
\end{equation}
where $F$ and $\mu$ are time independent constants defined by  
\begin{equation}
F \equiv \int_0^1 x P\ud{x}, 
\quad \mu \equiv \frac{\int_0^1 P\ud{x}}{\int_0^1 x^{\third} P\ud x}.\label{eq:F_and_mu}
\end{equation}
The solution of ODE \eqref{eq:c_dot} is  
\begin{equation}
  c(t') = \oneover{F\cdot(\mu-1)(t'-t_s)}\label{eq:c_of_t}.
\end{equation}
Here, $t_s$ is a time-shift related to the onset of
coarsening. It is determined later in the paper using the numerical solution
and knowledge of the similarity solution.

\begin{figure}[ht!]

  \centerline{
\resizebox{\columnwidth}{!}{\includegraphics{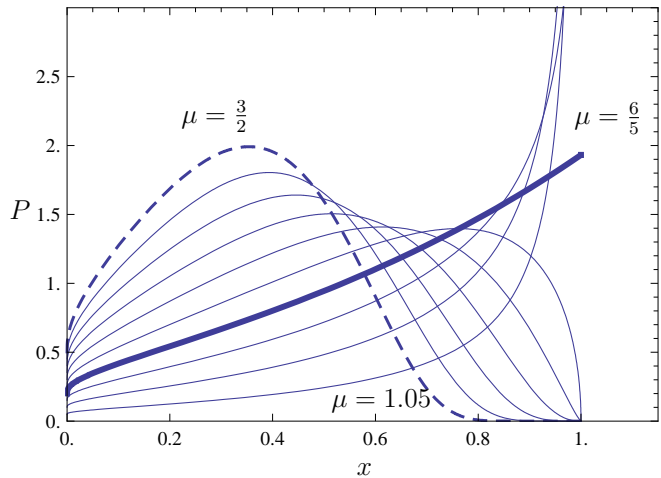}}}
\caption[Various similarity solutions]{Profiles $P(x)$ for  various values of
  $\mu$, $1<\mu\le\frac32$. 
The profiles are normalized so that $\int_0^1
P(x)\,dx=1$. Specifically, the values
of $\mu$ in the figure are $1.05$ through $1.5$ in steps of 0.05. 
The corresponding orders of contact vary from $-0.833$ through
0 (bold) and on to $\infty$ (dashed).
A function with order of contact $p\le-1$ is non-integrable, and
therefore unphysical.} 
\label{fig:similarity:sol}
\end{figure}

Given $c(t')$, we find the spatial part of the similarity solution, $P(x)$.
Substituting  \eqref{eq:ss:ansatz} into the convection PDE
\eqref{eq:PDE:coarsen:2}, and using ODE \eqref{eq:c_dot} for $c$ we
find an ODE for P:
\begin{equation}
  \frac{P_x}{P}= -\frac{\mu\brk{2-\third x^{-\frac23}}-2}{\mu(x^{\third}-x) + (x-1)}.\label{eq:ODE:P}
\end{equation}
Figure~\ref{fig:similarity:sol} shows $P(x)$  for different values of $\mu$.

\begin{figure}[ht!]
  \centerline{
\resizebox{\columnwidth}{!}{\includegraphics{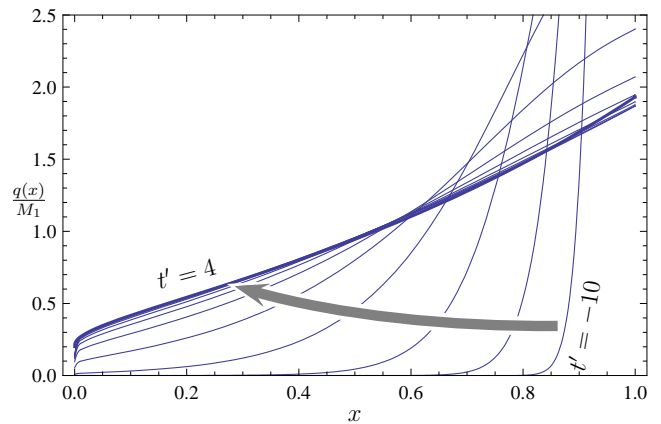}}}
\caption[The numerical solution approaching a similarity solution]{The numerical solution $q(x,t')$, scaled so that $M_1=1$, at
  different times. Starting on the right at $t'=-10$ in a narrow
  distribution, and very close to the similarity solution with \mbox{$\mu=\frac65$} (Dark
  line) at $t'=4$.}
\label{fig:converge}
\end{figure}

The parameter $\mu$ is related to the order of contact of $P(x)$ (with
zero) at $x=1$.
The order of contact is the power $p$ so that 
\begin{equation*}
  P(x)\sim b \,(1-x)^p \text{ as } x\goto 1^-,
\end{equation*}
for some constant $b>0$.
The super-script $(-)$ indicates that the limit is from below.
One sees that 
\begin{equation}
  p =\lim_{x\goto1^-} \tfrac{P_x}{P}(x-1).\label{eq:order:p}
\end{equation}
Substituting \eqref{eq:ODE:P} into \eqref{eq:order:p} gives
\begin{equation}
  p=\frac{5\mu-6}{3-2\mu}, \quad\text{or equivalently}\quad \mu = \frac{3p+6}{2p+5}.\label{eq:contact}
\end{equation}
Since the convection velocity $w$ in \eqref{eq:ad:vel:coarsen:2} is regular at $x=1$, the
order of contact of $q(x,\,t')$ at $x=1$ is constant, independent of
time \cite{NP99}.
The coarsening era solution is discontinuous at $x=1$,
so $p=0$, and then \eqref{eq:contact} implies $\mu=\frac65$.
We therefore \emph{expect} the numerical solution to
converge to the $\mu=\frac65$ similarity solution as $t\goto\infty$.
This convergence is verified numerically (see Fig.~\ref{fig:converge}.)

For  $\mu=\frac65$, the similarity solution profile $P(x)$ is:
\begin{equation}
P= \frac{125 \exp\!\!\brk{-\sqrt{\frac{12}{7}} \!\left(\coth^{\!\scriptscriptstyle -1}\!\!\left(\sqrt{21}\right)-\tanh^{\!\scriptscriptstyle-1}\!\!\left(\frac{2
   {x}^{1/3}+1}{\sqrt{21}}\right)\right)}}{\left(5-x^{2/3}-{x}^{1/3}\right)^3}.\label{eq:sim:sol:profile}
\end{equation}
Here, $P(x)$ is normalized so that
\begin{equation}
  \int_0^1 P \,dx=1.
\end{equation}
We estimate $F$ in \eqref{eq:F_and_mu} for this $P(x)$, numerically. It is
\newcommand{\F}{{0.632573}}
\begin{equation}
F\approx \F \label{eq:F}
\end{equation}
The dark line in Fig.~\ref{fig:similarity:sol} shows
$P$ with  $\mu=\frac65$. 
The dashed line shows the $c_\infty$ solution with $\mu=\frac32$.
The only parameter that is ``matched'' to the numerical data is the
time-shift, $t_s$, found via a simple method described next.
\begin{figure}[tb]
\centerline{
\resizebox{\columnwidth}{!}{\includegraphics{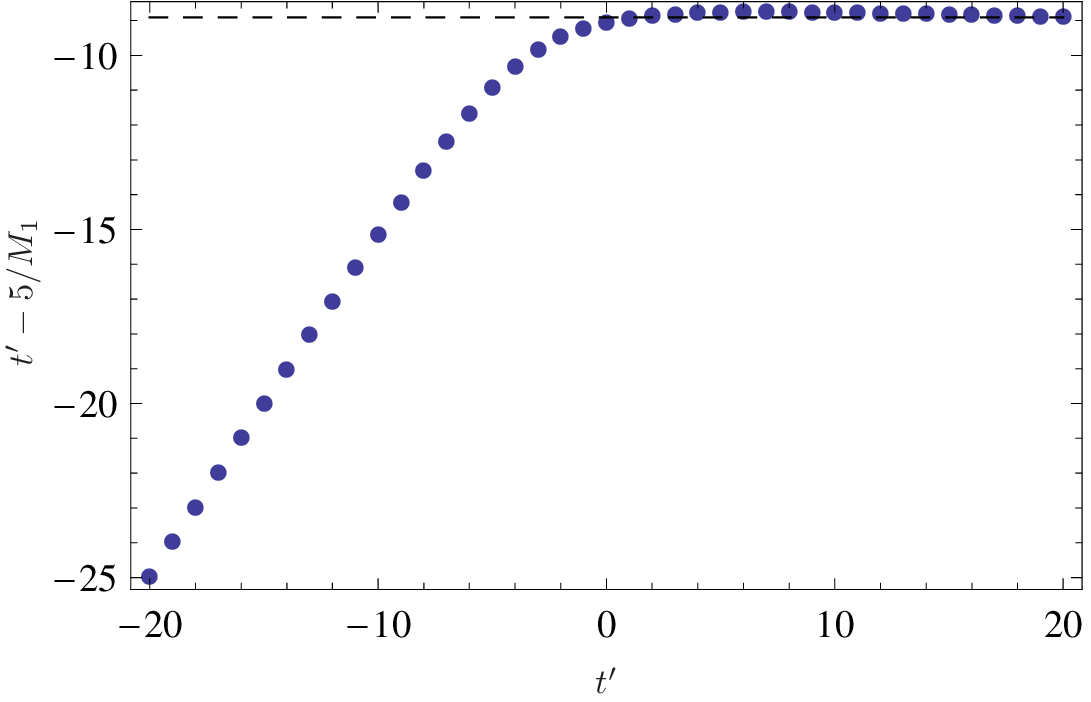}}}
\caption[Finding the time-shift for the similarity solution]{Finding
  the time-shift $t_s$. From around $t'=0$ and on the value of
  \mbox{$t'-\frac5{M_{1}(t')}$} stabilizes on $-\tsvalue$.}
\label{fig:finding_t_s}
\end{figure}

\subsection{Asymptotic Matching with Coarsening Era}
Finally, we determine the additive time constant $t_s$ in \eqref{eq:c_of_t} by 
examining the long-time limit of  the coarsening era solution. 
By substituting \mbox{$q(x,\,t')=c(t')P(x)$} with $c(t')$ as in \eqref{eq:c_of_t} into
\eqref{eq:Moments} for $M_1$, and setting $\mu=\frac65$, we find
\begin{equation}
  M_{1}(t') = \frac{5}{(t'-t_s)}\label{eq:M:t'}
\end{equation}
or equivalently,
\begin{equation}
 t_s = t'-\frac5{M(t')}.\label{eq:t_s}
\end{equation}
In order to estimate $t_s$, we calculate $M_{1}(t')$
from the numerical solution and plot the RHS of \eqref{eq:t_s} vs.\ $t'$ in Fig.~\ref{fig:finding_t_s}.
The horizontal asymptote as $t'\goto\infty$ shows that
$t_s\approx-\tsvalue$.
In Fig.~\ref{fig:NumericalMatching:f} we see that the numerical solution
indeed converges to the similarity solution with \mbox{$\mu=\frac65$} and
\mbox{$t_s=-\tsvalue$}.
From $t'=5$ and on the numerical and similarity solutions  are
practically indistinguishable. 
Thus, equations (\ref{eq:c_of_t}, \ref{eq:sim:sol:profile},
\ref{eq:F}) and the value of $t_s$ imply that the coarsening era solution, $q(x,\,t')$, asymptotes to 
\begin{equation}
\frac{625 \exp\!\brk{-2 \sqrt{\frac{3}{7}} \left(\coth ^{-1}\!\left(\sqrt{21}\right)-\tanh ^{-1}\!\left(\frac{2
   {x}^{1/3}+1}{\sqrt{21}}\right)\right)}}{\F \cdot(t'+\tsvalue)\left(5-x^{2/3}-{x}^{1/3}\right)^3}\label{eq:sim:sol:final}
\end{equation}
as $t\rightarrow\infty$.

\section{Methods}
\label{sec:methods}
Here we describe the method in which we solved the non-linear
convection PDE for the coarsening era.
We solve (\ref{eq:PDE:coarsen:2}--\ref{eq:eta:xq:1}) with the
initial condition \eqref{eq:effective:IC} using LeVeque's
conservation law numerical solver package, \texttt{clawpack}
\cite{LeVeque}, using the Riemann problem solver \texttt{rp1adecon} (a
Riemann solver for conservative convection). 
Since we expect the solution to start from a narrow and tall
distribution near $x=1$ and widen as $t$ increases, we use a
non-uniform grid that becomes more dense towards $x=1$. 
Specifically, 
\begin{equation}
x_n= \frac{11 n}{m+10n},
\end{equation}
where $m=2000$ is the number of grid-cells. 
Notice that $x_0=0$ and $x_m=1$.
This non-uniform grid is chosen so that it has a greater
resolution where we expect to find the biggest gradients, i.e.\ near $x=1$. 
The non-uniform grid was implemented using a variable ``capacity'' in the
numerical solver. 
The capacity of a cell denotes the change in mean value which results from a unit flux into the cell.
A non-uniform grid can be implemented on a uniform grid by giving the
computational cells that correspond to smaller physical cells a
smaller capacity, and the opposite for the larger cells, and adjusting
the convection velocity as needed. 
For more information on implementing non-uniform grid-size in \texttt{clawpack} please see \cite[Section 6.17]{LeVeque}.

We start the numerical solution at $t'=-20=3\log\delta$, so $\delta=\E^{-20/3}\approx1.2726\times10^{-3}$.
The integrals in \eqref{eq:Moments} are
calculated as Riemann sums\footnote{We use Riemann sums and not the
  trapezoidal rule. In \texttt{clawpack}, the cell values represent cell
  averages, and thus Riemann sum is the correct estimate. } 
By comparing the results to those obtained from a finer mesh and
smaller $\delta$, we estimate the relative error to be $\sim1\%$.

\section{Conclusion}
\label{sec:conclusion}
The original LS theory with its scale invariance describes late stage self-similar coarsening, but not the actual process of how it arises from the initial condition of pure monomer.
In particular, there is no prediction of characteristic cluster size, nor characteristic time that marks the onset of coarsening.
This paper, together with its predecessor on the creation era fills this gap in a limited sense:
It tells the story of the intermediate processes according to a ``classical'' aggregation kinetics based on a conservative union of fundamental ideas due to Becker-D\"oring, Zeldovich, and Lifshitz-Slyozov.
As we have seen, there is a succession of three eras, 
\emph{creation}, \emph{growth}, and \emph{coarsening}, with each consecutive era linked to the previous one by asymptotic matching.
In this paper, there is the additional twist of connecting analytic solutions to one universal numerical solution for coarsening.

The characteristic cluster size at the onset of coarsening is
\begin{equation}
[n]_{\text{c}}=\frac{\sig^3}{\eps^2R}[n], 
\end{equation}
where $[n]$ is the characteristic cluster size \eqref{eq:scale:n}
during creation, so the prefactor $\frac{\sig^3}{\eps^2R}$ is the
relative increase of size between creation and the onset of
coarsening.
The creation era size $[n]$ itself is dominated by the large exponential
$\exp\!\!\brk{\tfrac35 {G_*}/{k_BT}}$, so the free energy barrier against
nucleation $G_*$ is the most important physical parameter in the
determination of the characteristic size of $n_{\text{c}}$.

The time to onset of coarsening is a little more subtle than simple
scaling alone.
From (\ref{eq:N:M}, \ref{eq:M:t'}) the largest cluster size (in units of   $[n]_\text{c}$) is
\begin{equation}
N(t')=\tfrac{1}{5}(t'-t_s) \text{ for } t'\gg1.\label{eq:N:t'}
\end{equation}
The long time similarity solution in section~\ref{sec:coarsening} can be said to begin around $t'=0$. 
This is when the graph of $t'-\frac5M$ in Fig.~\ref{fig:finding_t_s}
becomes constant, as it should for the similarity solution, and when
the distance between the numerical solution and the similarity
solution is small and shrinking as can be seen in Fig.~\ref{fig:NumericalMatching:f}.
Recall that $t'$ is the shifted time as in \eqref{eq:t':t}, so equation
\eqref{eq:N:t'} reproduces the well-known result that the largest clusters
size grow linearly with time during late-stage coarsening.
The shifted time $t'$ is related to physical time $t$ by translation
and scaling, according to
\begin{equation}
t=[t]_\text{c} \brk{t'+2\log\tfrac{\sig^3}{\eps^2R}}.\label{eq:physical:t:coarsen}
\end{equation}
Here, $[t]_\text{c}$ is the scaling unit of time from the
coarsening era,
\begin{equation}
[t]_{\text{c}}=\tfrac{N_0}{s}[t]\label{eq:coarsen:t:scale}
\end{equation}
or, using $s$ as in \eqref{eq:s} and $[t]$ as in \eqref{eq:scale:t}, 
\begin{equation}
[t]_{\text{c}}=\tfrac{\sig^3}{\eps^2}[n]\brk{\tfrac{1}{Dv^\third f_s}}.
\end{equation}
The factor $1/{Dv^\third f_s}$ carries the unit of time.
Notice that $[t]_{\text{c}}$, like $[n]_{\text{c}}$, is
proportional to $\exp\!\brk{\tfrac35 {G_*}/{k_BT}}$.
This is exponentially longer than the creation timescale $[t]$ which is proportional to $\exp\!\brk{\tfrac25{G_*}/{k_BT}}$.
The additive time constant $2\log\frac{\sig^3}{\eps^2R}$ in \eqref{eq:physical:t:coarsen} comes from \eqref{eq:t':t} with $N_0=\frac{\sig^3}{\eps^2R}$.
Setting $t'=0$ in \eqref{eq:physical:t:coarsen} we recover the onset
of late stage  coarsening in physical time,
\begin{equation}
t_\text{onset}=2 \,[t]_\text{c} \log\tfrac{\sig^3}{\eps^2R}.
\end{equation}
As $\eps\goto0$ the logarithm term is nominally large.
But in the ``big picture'', it is just another prefactor which is
overwhelmed by the large exponential $\exp\!\brk{\tfrac35 {G_*}/{k_BT}}$ in
$[t]_\text{c}$.
In summary, the characteristic cluster size and characteristic time
associated with the onset of coarsening are both proportional to
$\exp\!\brk{\tfrac35{G_*}/{k_BT}}$.

\section{Discussion}
\label{sec:discussion}
Although we have described the three eras, as we have set out to do,
the aggregation story is not over.
According to conventional wisdom, the ``correct'' similarity solution
is the smooth one with $p=\infty$ (and $\mu=\frac32$) and not the discontinuous one we found with $p=0$.
In the original LS paper, it is suggested that the (rare) coagulation
of large clusters eventually selects the smooth similarity solution.
A recent work by Niethammer and Velasquez \cite{NV06} suggests that
screening-induced fluctuations also leads to the selection of the
smooth solution.

Even classical kinetics without any additional physics has
structure that is not yet fully examined. 
It can lead to the selection of the smooth solution given
 sufficiently long time, much longer than $t_\text{c}$ in
\eqref{eq:coarsen:t:scale}.
In particular, the convection PDE boundary value problem for $r(n,t)$
is the lowest order of approximation to a discrete system of ODE's. 
The next order of approximation introduces an effective diffusion in
size space, and this smooths out the discontinuity. 
Another effect of the discrete kinetics is that the Zeldovich
nucleation rate does not abruptly ``turn on'' at $t=0$ as we have
assumed in our reduced analysis.
It has been shown (both analytically \cite{NBC05} and experimentally \cite{KGT83}) that there is
a transient during which the nucleation rate \emph{smoothly}
increases, with the Zeldovich rate as its long time limit.  
This is the so-called ``ignition transient''.
Taking this transient into consideration would add a narrow boundary
layer to the sharp front, connecting it smoothly to zero. 
A preliminary analysis shows that a narrow boundary layer about $n=N$
widens on a timescale much longer than $[t]_\text{c}$, and the
smooth solution results.

In summary, the discontinuous similarity solution is structurally
unstable due to a variety of physical and mathematical
perturbations. 
It is now a question of time scales: 
The mechanism that causes the fastest deviation from the
discontinuous solution will determine the timescale of this last era, and will be the main cause for the smoothing of the distribution.
This ``contest'' will be played out in a future paper.

\appendix
\renewcommand{\theequation}{\Alph{section}.\arabic{equation}}

\section{Coarsening Era Scaling}
\label{app:scaling}
The scales $[t], [r],$ and $[n]$ of time, cluster density and cluster size of the creation era are found (in \cite{FN:Creation:08}) to be
\begin{align}
 \Brk{t}&=(8\pi)^{-\oneover{5}}\BRK{\eps^\frac35\sig^{-\frac75}}\E^{\frac25\frac{G_*}{k_BT}}\brk{D^3vf_s^3\omega^2}^{-\frac15},
  \label{eq:scale:t}\\
\Brk{n}&=(\pi^\frac{7}{10}2^\frac{11}{10}\sqrt{3})\BRK{\frac{D \eps^4 f_s v^{\frac13}}{\sig^{\frac{7}{2}}\omega}}^{\frac35}\E^{\frac35\frac{G_*}{k_BT}},\label{eq:scale:n}\\
\Brk{r}&=(3\cdot 2^{11}\pi^7)^{-1/5}\BRK{\frac{\sig^2\omega^2 }{\eps^3 D^2f_s^2v^{\frac23}}}^{\frac35}\E^{-\frac65\frac{G_*}{k_BT}}(f_s)\label{eq:scale:r}.
\end{align}
Here, $\eps$ denotes the initial (small) value of the super-saturation $\eta(0)$.
In the exponents, the fraction $\frac{\sig^{3}}{2\eta^{2}}$ approximates the initial free energy barrier against nucleation in units of $k_{B}T$. 
In the prefactors, $\omega$ is the evaporation rate so that $\omega n^{\frac23}$ is the rate at which monomers at the surface of an $n-$cluster leave it.
The dominant balances leading to these scaling units are based physically upon the Zeldovich rate of nucleation, diffusion limited growth of created clusters, and conservation of monomers. 
In particular the exponential largeness of characteristic time and cluster size $[t]$ and $[n]$ in $\eps$ arise from the exponential smallness of the Zeldovich nucleation rate (proportional to $\exp-{\sig^{3}}/{2\eta^{3}}$).
The relation $[n]\propto[t]^{\frac32}$ as evident from
(\ref{eq:scale:t}, \ref{eq:scale:n}) is a signature of diffusion
limited growth (in 3D).

The scaled surface tension $s$ in the scaled version of the LS equation \eqref{eq:PDE:scaled} is given by
\begin{equation}
s=\brk{3(4\pi)^{2}}^{\third}(Dv^{\third}f_{s})\tfrac{[t]}{[n]}\sig.
\label{eq:s}
\end{equation}
From~(\ref{eq:scale:t}, \ref{eq:scale:n}) we see that $s\propto
\exp\!\brk{-\tfrac15{\sig^{3}}/{2\eps^{2}}}$ is exponentially small for $\eps\ll1$.
\section{Growth Era Solution}
\label{app:growth}
Let us, for the moment, take $\eta(t)$ as given.
The characteristic curve corresponding to the ``first'' cluster---the
one that nucleated at
time $t=0$---is $n=N(t)$, where $N(t)$ satisfies
\begin{equation}
\dot N = \eta N^\third, N(0) = 0,\, N(t)>0 \text{ for } t>0.
\label{eq:ODE:N:growth}
\end{equation}
The support of $r(n,\,t)$ lies in $\RR$:
\begin{equation}
\RR \equiv \BRK{ (n,\,t) : 0 < n < N(t),\, t>0}.
\end{equation}
In $\RR$ the value of $r(n,t)$ is found from 
\begin{equation}
r(n,\,t) = n^{-\third}g(\tau),\label{eq:r:tau}
\end{equation}
where $g(\tau)$ is the constant value of $n^\third r(n,\,t)$ along the
characteristic curve that has $n(\tau)=0$.
For any point in $\RR$, there is one characteristic curve that passes
through it, so $\tau$ in \eqref{eq:r:tau} is a function of $n$ and $t$.
Given $\tau=\tau(n,\,t)$, \eqref{eq:r:tau} is the growth era solution for
$r(n,\,t)$ in $\RR$.

The asymptotic determinations of $g(\tau)$ and $\tau(n,\,t)$ are simple.
It follows from~(\ref{eq:char:n}, \ref{eq:ODE:N:growth}) that
\begin{align}
\tfrac32N(t)^{\frac23} &= \int_0^t \eta(t')dt',\\
\tfrac32n^{\frac23} &= \int_{\tau(n,\,t)}^t \eta(t')dt'.
\intertext{Subtracting these equations gives}
\tfrac32\brk{N^{\frac23}-n^{\frac23}} &= \int_0^{\tau(n,\,t)} \eta(t')dt'.\label{eq:diff:n:N}
\end{align}
Characteristics with $\tau=\O(1)$ are launched during the creation
era, and for these we have  that $g(\tau)$ in \eqref{eq:r:tau} is
in fact $j(\tau)$ from \eqref{eq:nucleation:flux}.
During creation, $\eta(t)$ (in units of $\eps$) differs from 1 by
$\O(\eps^2)$, so for $\tau=\O(1)$, we replace $\eta(t')$ in \eqref{eq:diff:n:N}
by 1, 
\begin{equation}
\tau =\tfrac32\brk{N^\frac23-n^\frac23}.\label{eq:tau:formula}
\end{equation}
In the limit $N\gg1$, the RHS of \eqref{eq:tau:formula} remains $\O(1)$ for $N-n=\O(N^\third)$
and in this case we replace the RHS by its linearization about $n=N$,
so
\begin{equation}
\tau \sim \tfrac{N-n}{N^\third}.\label{eq:tau}
\end{equation}
Once we determine $N=N(t)$, \eqref{eq:tau} gives $\tau(n,\,t)$ and the
solution for $r(n,\,t)$ in the ``front''
\begin{equation}
0<N(t)-n = \O\brk{N(t)}^\third.\label{eq:narrow:front}
\end{equation}
is given by
\begin{equation}
r(n,\,t) \sim N^{-\third} j\brk{\tfrac{N-n}{N^\third}}.
\end{equation}
For $N-n\gg N^\third$, $\tau\gg1$ and $j(\tau)$ asymptotes to zero,
corresponding to negligible production of new clusters \emph{after}
the creation era.

We complete the story of the growth era by an asymptotic determination
of $N(t)$.
Since the support of $r(n,\,t)$ is effectively the narrow front
\eqref{eq:narrow:front}, the conservation identity
\eqref{eq:conservation:scaled} reduces asymptotically to
\begin{equation}
\eta(t) \sim 1-\tfrac{\eps^2 R}{\sig^3} N.
\label{eq:conserv:asympt:growth}
\end{equation}
Here, $R$ is the (scaled) total number of clusters produced during
nucleation, given by \eqref{eq:total:density:nuc}.
Combining~(\ref{eq:ODE:N:growth}, \ref{eq:conserv:asympt:growth})
gives a simple ODE for $N(t)$,
\begin{equation}
\dot N \sim \brk{1- \tfrac{N}{N_0}} N^\third,
\label{eq:ODE:N:growth:2}
\end{equation}
where
\begin{equation}
N_0 \equiv \tfrac{\sig^3}{\eps^2 R}.
\end{equation}
The solution with $N(0)=0$ (and $N>0$ for $t>0$) is given implicitly by
\begin{equation}
\frac{t}{N_0^\frac23} = \sum_{j=0}^2 r_j
\log\brk{1+r_j\brk{\tfrac{N}{N_0}}^\third}.
\end{equation}
Here, $r_j$ are the cube roots of $-1$: $r_0=\E^{i\frac{\pi}3}, r_1=-1,
r_2=\E^{-i\frac{\pi}3}$.

\bibliography{general}

\end{document}